\documentstyle[aps,floats]{revtex}

\begin{document}
\input epsf
\renewcommand{\topfraction}{1.0}
\twocolumn[\hsize\textwidth\columnwidth\hsize\csname@twocolumnfalse\endcsname

\title{New limits on dark--matter WIMPs from the Heidelberg--Moscow experiment}
\author{L. Baudis, J. Hellmig, G. Heusser, H.V. Klapdor--Kleingrothaus
\cite{sp}, 
S. Kolb, B. Majorovits, H. P\"as, Y. Ramachers and H. Strecker}
\address{Max--Planck--Institut f\"ur Kernphysik,
P.O.Box 10 39 80, D--69029 Heidelberg, Germany}
\author{V. Alexeev, A. Bakalyarov, A. Balysh, S.T. Belyaev
\cite{sp}, V.I. Lebedev and S. Zhukov}
\address{Russian Science Centre Kurchatov Institute, 123 182 Moscow, Russia}

\maketitle
\begin{abstract}
New results after 0.69 kg$\,$yr of measurement with an 
enriched $^{76}$Ge detector 
of the Heidelberg--Moscow experiment with an active mass of 2.758 kg are presented. 
An energy threshold of 9 keV and a background level of 
0.042 counts/(kg$\,$d$\,$keV) in the energy 
region between 15 keV and 40 keV was reached.
The derived limits on the WIMP--nucleon cross section are 
the most stringent limits on spin--independent interactions 
obtained to date by using essentially raw data without 
background subtraction.
\noindent {\it PACS number(s):} 95.35.+d, 14.80.Ly
\end{abstract}
\vskip2pc]

\section*{Introduction}

The nature of dark matter in the Universe remains a challenging question. 
Even if new measurements will confirm that we live in a low 
$\Omega_{\text{matter}}$ universe ($\Omega_{\text{matter}}\sim 0.3 - 0.4 $) 
\cite{perlm98,riess98} a considerable amount of nonbaryonic 
dark matter is needed.  
WIMPs (weakly interacting massive particles) are among the most 
discussed candidates
\cite{jung96}, being well motivated from early universe physics 
\cite{kolb94} and supersymmetry \cite{hab85}.

WIMP detection experiments can decide whether WIMPs 
dominate the halo of our Galaxy. For this reason, considerable
effort is made towards direct WIMP search experiments which look
for energy depositions from elastic WIMP--nucleus scattering 
\cite{smith90}. Germanium experiments designed for the search for neutrinoless
double beta decay were among the first to set such kind of limits
\cite{ahlen87,caldwell88}. 

The Heidelberg--Moscow experiment gave the most stringent upper limits on 
spin--independent WIMP interactions \cite{beck94} until recently. 
The present best limits on the WIMP--nucleon cross section 
come from the DAMA NaI Experiment \cite{bern97}.  
The Heidelberg--Moscow experiment operates five enriched 
$^{76}$Ge detectors with an active mass of 10.96 kg in the Gran 
Sasso Underground Laboratory. It is optimized for the search for the
neutrinoless double beta decay of $^{76}$Ge in the energy region of 2038 
keV. For a detailed description of the experiment 
and latest results see \cite{heidmo}.

In this paper we report on results from one of the enriched
Ge detectors, which took data in a period of 0.249 years
in a special configuration developed for low energy measurements. 
A lower energy threshold and a smaller background counting rate 
has been achieved with the same detector as used in 1994 
\cite{beck94}, mainly due to the lower cosmogenic activities in the 
Ge crystal and in the surrounding copper after four years without 
activation. 

\section*{Experimental setup}

The utilized detector is a coaxial, intrinsic p--type HPGe detector 
with an active mass of 2.758 kg. The enrichment in $^{76}$Ge is
86\%. The sensitivity to spin--dependent interactions 
becomes negligible, since $^{73}$Ge, the only stable Ge isotope 
with nonzero spin, is deenriched to 0.12\% (7.8\% for natural Ge).
The detector has been in the Gran Sasso Underground Laboratory since 
September 1991; a detailed description of its background can be found in 
\cite{heidmo}. 

The data acquisition system allows an event-by-event 
sampling and pulse shape measurements. 
The energy output of the preamplifier is divided and amplified with two
different shaping time constants, 2 $\mu$s and 4 $\mu$s. The fast 2 
$\mu$s signal
serves as a stop signal for the 250 MHz flash analogue to digital
converter (FADC) which records the 
pulse shape of each interaction. 
The best energy resolution and thus lowest energy 
threshold is obtained with the 4 $\mu$s shaped signal.
A third branch of the energy output is shaped with 3 $\mu$s and 
amplified to record
events up to 8 MeV in order to identify radioactive impurities
contributing to the background. The spectra are measured with 13bit ADCs, 
which also release the
trigger for an event by a peak detect signal.
Further triggers are vetoed until the 
complete event information, including the pulse shape, has been recorded.
To record the pulse shape (for details see \cite{laura-nim})
the timing output of the preamplifier is 
divided into four branches, each signal being integrated and 
differentiated in filtering amplifiers (TFAs) with different time 
constants. The TFAs are used since the charge current is 
integrated within the preamplifier. The signals are amplified 
to record low as well as high--energetic pulses. 

\section*{Data analysis}

An energy threshold of 9 keV has been 
reached. This rather high value is due to the large detector size and 
a 50 cm distance between FET and detector. Both effects lead to a higher 
system capacitance and thus to an enhancement of the baseline 
noise.

We calibrate the detector with a standard $^{152}$Eu--$^{228}$Th source. 
The energy resolution at 
727 keV is (2.37 $\pm$ 0.01) keV. In order to determine the energy 
resolution at 0 keV 
the dependence of the full width at half maximum (FWHM) on the energy
is approx imated with an empirical function 
$y=\sqrt{a + b\,x + c\,x^2}$ (y = resolution, x = energy) in a 
$\chi ^2$ fit. The best fit ($\chi ^2$/DOF = 0.09) is
obtained for the parameters a = 3.8, b = 2.2$\times 10^{-3}$, c = 5$\times 10^{-7}$. 
The zero energy 
resolution is (2 $\pm$ 0.01) keV.

To determine the energy threshold, a reliable energy calibration at low
energies is required. The lowest energetic line observed in the 
detector was at 74.97 keV (K$_\alpha$ line of $^{208}$Pb).
 This is due to the rather thick copper shield of the 
crystal (2 mm) which absorbs all low energy $\gamma$ lines.
Thus an extrapolation of the energy calibration to low energies is 
needed, which induces an error of (1--2) keV.
Another possibility is to use a precision pulser in order to 
determine the channel of the energy spectrum which corresponds to zero 
voltage pulser output. Since the slope of the energy calibration is 
independent of the intercept and can be determined with a calibration 
source, this method yields an accurate value for the intercept 
of the calibration. 
The same method to determine the offset of the calibration is also used
by \cite{beck94} and \cite{reusser91}. The pulser calibration reduces 
the extrapolated 9 keV threshold systematically by (1--2) keV. In order to 
give conservative results, we use the 9 keV threshold for data analysis.

\begin{figure}
\label{burst}
\centering
\leavevmode\epsfysize=7cm\epsfbox{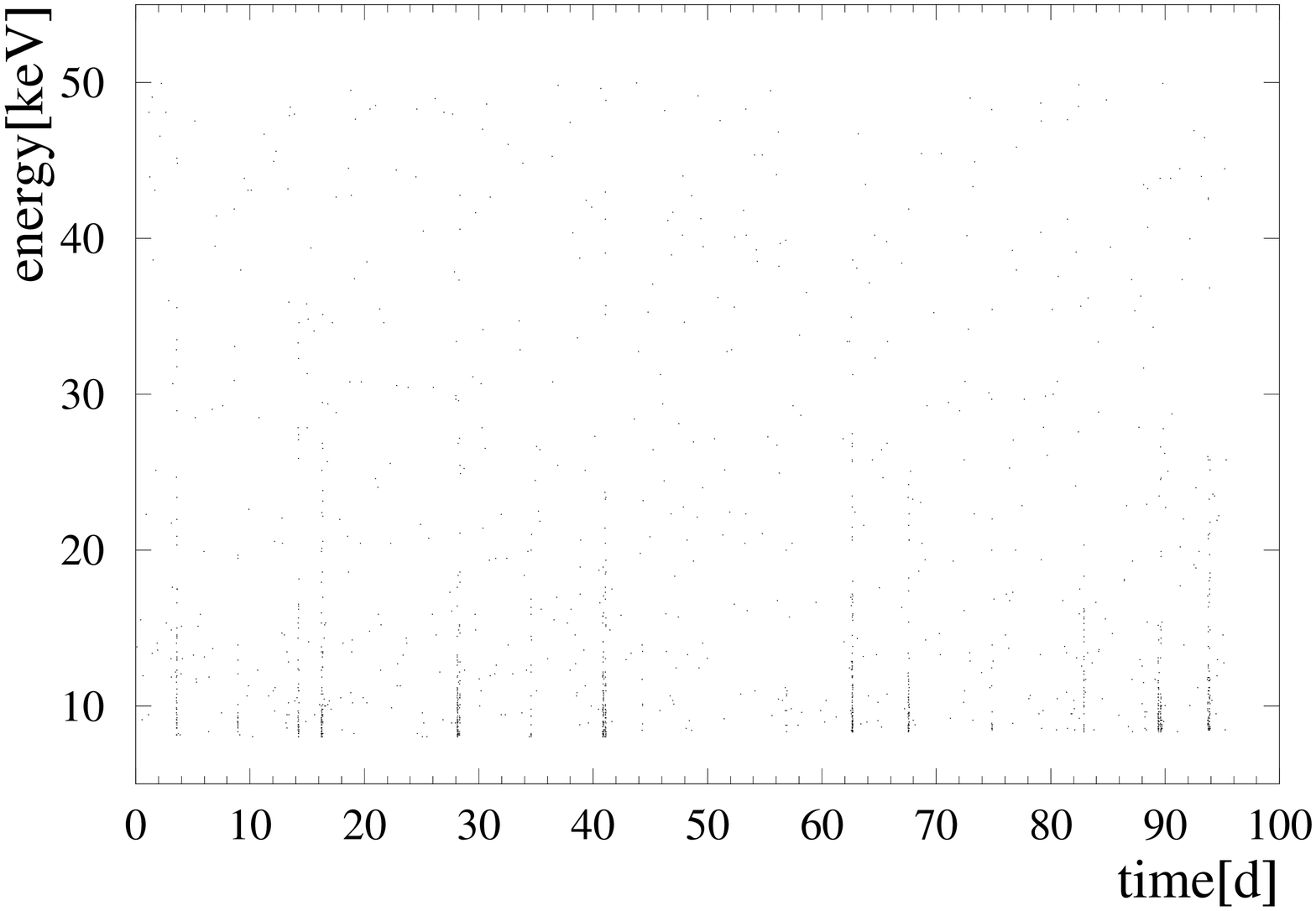}
\caption{Energy spectrum as a function of time for the 
enriched $^{76}$Ge detector for the whole measurement period. Irregular 
bursts up to 30 keV can be seen.}
\end{figure}

In Fig.~\ref{burst} the energy deposition as a function of time is 
plotted. Accumulation of events (bursts) with energy depositions up to 
30 keV can be seen. They are irregularly distributed in time with 
counting rate excesses up to a factor of five. 
These events can be generated 
by microphonics (small changes in the detector capacitance due to 
changes in temperature or mechanical vibrations) or by a raised 
electronic noise on the baseline. 
Although rare, they lead to an enhancement
of the count rate in the low energy region. A possibility to deal 
with microphonics would be to exclude 
the few days with an enhanced count rate from the sum spectrum like 
in \cite{ahlen87}. This would lead, however, to unwanted measuring time 
losses. Another time filtering method is applied in the following. 

\begin{figure}
\label{poisson}
\centering
\leavevmode\epsfysize=7cm\epsfbox{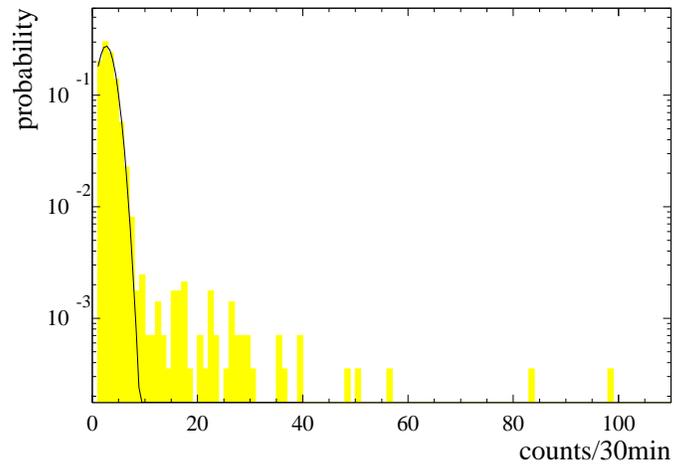}
\caption{Distribution of the number of events with energies between 9 keV 
and 850 keV per 30--minute interval. The solid curve represents the 
Poisson fit to the data.}
\end{figure}

The complete measuring time is divided into 30-minute intervals and 
the probability for N events to occur in one time interval is 
computed for energy depositions between 9--850 keV. In the histogram in 
Fig.~\ref{poisson} the physical events are Poisson distributed. The mean 
value of the distribution is (2.65 $\pm$ 0.06) events/30 min and 
$\sigma$=(1.67$\pm$ 0.05). The cut is set at the value N = 2.65 + 
3$\,\sigma \approx $ 8. With this cut less than 0.01\% of the 
undisturbed 30 minutes intervals are rejected. The initial 
exposure  of the measurement was 0.7 kg$\,$yr, 
after the time cut the exposure is 0.69 kg$\,$yr. In this 
way, more than 98\% of the initial data are used. A similar method to reduce 
the microphonic noise in the low energy region was also applied by \cite{beck94}.

Another way to reject microphonic events would be to analyze the 
recorded pulse shapes of each event. In a former experiment we 
showed \cite{laura-nim} that pulse shapes of nuclear recoil and 
$\gamma$ interactions are indistinguishable within the timing 
resolution of Ge detectors. Thus a reduction of $\gamma$--ray 
background based on pulse shape discrimination (PSD) is not possible. 
Consequently $\gamma$ sources can be employed to calibrate a PSD 
method against microphonics, which was shown to reveal a different 
pattern in the pulse shape. Since such a pulse shape analysing method 
is still under development, we use only the Poisson--time--cut method 
in this paper.

\begin{figure}
\label{sumspec}
\centering
\leavevmode\epsfysize=11cm\epsfbox{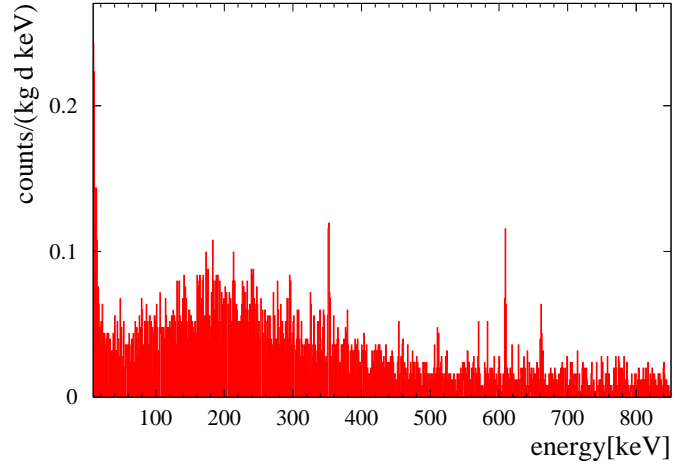}
\caption{Full recorded sum spectrum with lifetime 0.69 kg$\,$yr. Peaks from 
the $^{238}$U chain can be identified. A $^{137}$Cs peak slowly appears 
at 662 keV. All structures at low energies are fluctuations so far and 
therefore not identifiable.}
\end{figure}

Figure \ref{sumspec} shows the sum spectrum after the time cut. 
The background counting rate in the energy region between 9 keV 
and 30 keV is 0.081 cts/(kg$\,$d$\,$keV) [between 15 keV and 40 keV: 
0.042 cts/(kg$\,$d$\,$keV)]. This is about a factor of
two better than the background level reached by \cite{beck94} with 
the same Ge detector. Table~\ref{table1} gives the number of counts 
per 1 keV bin for the energy region between (9--50) keV. The dominating 
background contribution in the low--energy region from the U/Th natural 
decay chain can be identified in Fig.~\ref{sumspec} via the 352 keV and 
609 keV lines (the continuous beta spectrum from $^{210}$Bi originates from this chain).

\section*{Dark Matter Limits}

The evaluation for dark matter limits on the WIMP--nucleon cross section 
$\sigma_{\rm scalar}^{\rm W-N}$ follows the conservative 
assumption that the whole experimental spectrum consists of 
WIMP events.
Consequently, excess events from calculated 
WIMP spectra above the experimental spectrum in any energy region 
with a minimum width of the energy resolution of the detector 
are forbidden (to a given confidence limit).

The parameters used in the calculation of expected WIMP spectra are 
summarized in Table ~\ref{tab:parameter}. We use formulas given 
in the extensive reviews \cite{rita,lewin} for a truncated 
Maxwell velocity distribution in an isothermal WIMP--halo 
model (truncation at the escape velocity, compare also \cite{freese}).

Since $^{76}$Ge is a spin zero nucleus, we give cross section limits for 
WIMP--nucleon scalar interactions only. For these, we used the 
Bessel form factor (see \cite{lewin} and references therein) for the 
parametrization of coherence loss, adopting a skin thickness of 1 fm. 

Another correction which has to be applied for a semiconductor 
ionization detector is the ionization efficiency. There exist 
analytic expressions \cite{smith90} for this efficiency, especially for 
germanium detectors and multiple experimental results measuring this 
quantity (see \cite{laura-nim} and references therein). According to our 
measurements \cite{laura-nim} we give a simple relation between visible 
energies and recoil energies: 
$E_{\text{vis}} = 0.14\,E_{\text{recoil}}^{1.19}$. 
This relation has been checked for consistency with the relation 
from \cite{smith90} in the relevant low energy region above our threshold. 

\begin{figure}
\label{spectrum-wimp}
\centering
\leavevmode\epsfysize=11cm\epsfbox{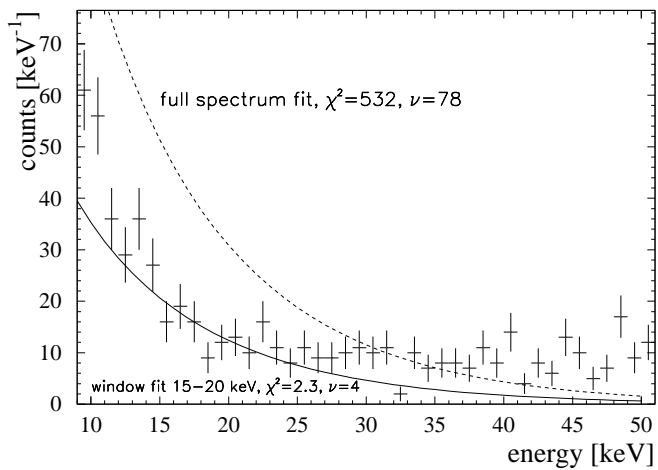}
\caption{Comparison of the measured low energy spectrum (shown from 
threshold to 50 keV) from the 
enriched $^{76}$Ge Detector and calculated WIMP spectra for a 100 GeV 
WIMP, already fitted for the allowed cross section 
$\sigma_{\text{scalar}}^{\text{W-N}}$. The solid curve shows the result 
from the sliding--window fit (see text). The dashed curve would result 
for a full WIMP--spectrum fit to the data, yielding too conservative 
limits.}
\end{figure}

After calculating the WIMP spectrum for a given WIMP mass, the scalar 
cross section is the only free parameter which is then used to fit the 
expected to the measured spectrum (see Fig.~\ref{spectrum-wimp}) 
using a one--parameter 
maximum--likelihood fit algorithm. According to the underlying 
hypothesis (see above) we check during the fit for excess events above 
the experimental spectrum (for a one--sided 90\% C.L.) using a sliding, 
variable energy window. The minimum width of this energy window 
is 5 keV, corresponding to 2.5 times the FWHM of the 
detector (6$\sigma$ width). 
The minimum of cross section values obtained via these multiple fits of 
the expected to the measured spectrum gives the limit.
Figure ~\ref{spectrum-wimp} shows a comparison between the measured
spectrum and the calculated WIMP spectrum for a 100 GeV WIMP mass.
The solid curve represents the fitted WIMP spectrum using a minimum width of
5 keV for the energy window. The minimum is found in the energy region
between 15 keV and 20 keV.
The dashed line is the result of the fit
if the energy window width equals the full spectrum width. It is easy to see
that in this case the obtained limit would be much too conservative,
leading to a loss of the information one gets from the measured
spectrum.

\section*{Conclusions}

\begin{figure}
\label{dm_limits}
\centering
\leavevmode\epsfysize=7cm\epsfbox{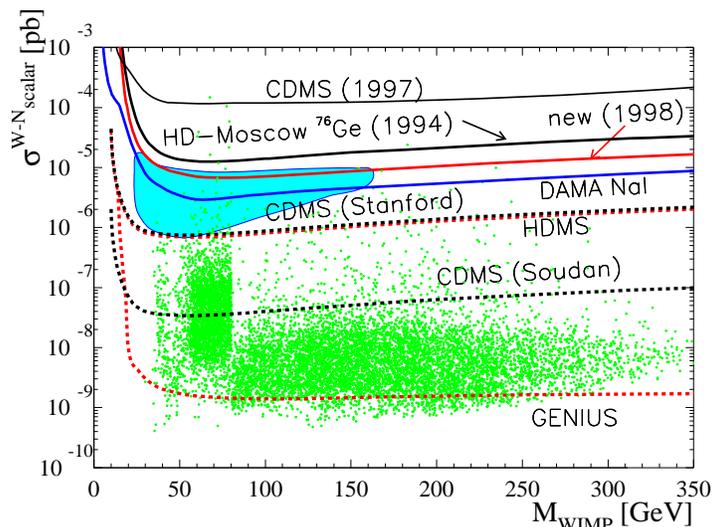}
\caption{Comparison of already achieved WIMP--nucleon scalar cross section
limits (solid lines): the Heidelberg--Moscow $^{76}$Ge 
\protect{\cite{beck94}}, the
recent CDMS nat. Ge \protect{\cite{cdms98}} and the new DAMA NaI result
\protect{\cite{bern97}}, including their evidence contour 
\protect{\cite{DAMA}}, in pb for scalar interactions as
function of the WIMP mass in 
GeV and of possible results from upcoming experiments (dashed lines for
HDMS \protect{\cite{hdms97}}, CDMS (at different locations) and GENIUS 
\protect{\cite{genius97}}). 
These experimental limits are also compared to
expectations (scatter plot) for WIMP neutralinos calculated in the
MSSM framework with non--universal scalar mass unification
\protect{\cite{bedny1}} (all curves and dots scaled for $\rho=0.5$ 
GeV/cm$^{3}$ for comparison of published data).}
\end{figure}

The new upper limit exclusion plot in the 
$\sigma_{\text{scalar}}^{\text{W-N}}$ versus M$_{\text{WIMP}}$
plane is shown in Fig.~\ref{dm_limits}. Since we do not use any
background subtraction in our analysis, we consider our limit to be 
conservative. 
We are now sensitive to WIMP masses greater than 13 GeV and to
cross sections as low as 1.12$\times$10$^{-5}$ pb 
(for $\rho=0.3$ GeV/cm$^{3}$).

At the same time we start to enter the region (evidence--contour)
allowed with 90\% C.L. if the preliminary analysis of   
4549 kg days of data by the DAMA NaI Experiment \cite{DAMA}
are interpreted as an evidence for an annual modulation 
effect due to a spin independent coupled WIMP. 

Should the effect be confirmed with much higher statistics
(20 000 kg days are now being analyzed by the DAMA Collaboration
\cite{rita98}) it could become crucial to test the region
using a different detector technique and a different target material.

Also shown in the figure are recent limits from the CDMS Experiment
\cite{cdms98}, from the DAMA Experiment \cite{bern97}, 
as well as expectations for new dark matter experiments like
CDMS \cite{cdms98}, HDMS \cite{hdms97} and for our recently proposed 
experiment GENIUS \cite{genius97}.
Not shown is the limit from the UKDM Experiment \cite{ukdm}
which lies somewhere between the two germanium limits.

After a measuring period of 0.69 kg$\,$yr with one of the enriched 
germanium detectors of the Heidelberg--Moscow experiment,
the background level decreased to 0.0419 
counts/(kg$\,$d$\,$keV) in the low energy region.
The WIMP--nucleon cross section limits for spin--independent
interactions are the most stringent limits obtained 
so far by using  essentially raw data without  background subtraction. 

An improvement in sensitivity could be reached after a longer measuring
period. Higher statistics would allow the identification of the various 
radioactive background sources and would open the possibility of a
quantitative and model--independent background description via a Monte 
Carlo simulation. Such a background model has already been established
for three of the enriched Ge detectors in the Heidelberg--Moscow
experiment
and has been successfully applied in the evaluation of the
2$\nu\beta\beta$ decay \cite{heidmo}. A subtraction of background 
in the low energy region in the form of a phenomenological straight line
based on a quantitative background model for the full energy region
(9 keV -- 8 MeV) would lead to a further improvement 
in sensitivity. Background
subtractions for dark matter evaluations of Ge experiments were
already applied by \cite{garcia}.

Another way to reject radioactive background originating from multiple 
scattered photons would be an active shielding in the immediate
vicinity of the measuring crystal. This method is applied in our
new dark matter search experiment, HDMS (Heidelberg Dark Matter
Search) \cite{hdms97}. HDMS is situated in the Gran Sasso Underground
Laboratory and started operation this year.  
 
\acknowledgments 
The Heidelberg--Moscow experiment was supported by the 
Bundesministerium f\"ur Forschung und Technologie der Bundesrepublik 
Deutschland, the State Committee of Atomic Energy of Russia and the 
Istituto Nazionale di Fisica Nucleare of Italy. L.B. was supported by 
the Graduiertenkolleg of the University of Heidelberg.

\onecolumn

\begin{table}
\caption{Number of counts per 1 keV energy bin after 250.836 kg$\,$d.}
\label{table1}
\begin{tabular}{rcrc}
Bin [keV]&Counts [1/keV]&Bin [keV]&Counts [1/keV]\\\hline
9-10 &61 &30-31 &10 \\
10-11 &56 &31-32 &11 \\
11-12 &36 &32-33 &2 \\
12-13 &29 &33-34 &10 \\
13-14 &36 &34-35 &7 \\
14-15 &27 &35-36 &8 \\
15-16 &16 &36-37 &8 \\
16-17 &19 &37-38 &7 \\
17-18 &16 &38-39 &11 \\
18-19 &9 &39-40 &8 \\
19-20 &12 &40-41 &14 \\
20-21 &13 &41-42 &4 \\
21-22 &10 &42-43 &8 \\
22-23 &16 &43-44 &6 \\
23-24 &11 &44-45 &13 \\
24-25 &8 &45-46 &10 \\
25-26 &11 &46-47 &5 \\
26-27 &9 &47-48 &7 \\
27-28 &9 &48-49 &17 \\
28-29 &10 &49-50 &9 \\
29-30 &11 &50-51 &12 \\
\end{tabular}
\end{table}

\begin{table}[htb]
\caption{List of parameters used for calculating WIMP spectra.}
\begin{tabular}{ll}
Parameter&Value\\\hline
WIMP velocity distribution& 270 km/s\\
Escape velocity& 580 km/s\\
Earth velocity& 245 km/s\\
WIMP local density& 0.3 GeV/cm$^{3}$\\
enriched $^{76}$Ge mass& 75.14 g/mol\\
\end{tabular}
\label{tab:parameter}
\end{table}

\end{document}